\documentstyle[aps,prl,twocolumn,epsf,rotate]{revtex}

\newcommand{\kk}{{\bf k}}

\newcommand{\qq}{{\bf q}}
\newcommand{\QQ}{{\bf Q}}

\newcommand{\dd}{{\vec{\bf \delta}}}
\newcommand{\cc}{{\hat c}}
\newcommand{\ah}{{\hat a}}

\newcommand{\ccd}{{\hat c^\dagger}}

\begin{document}
\twocolumn[\hsize\textwidth\columnwidth\hsize\csname
@twocolumnfalse\endcsname

\title{Electron-Phonon Interactions in Polyacene Organic Transistors}

\draft
\author{Jairo Sinova, 
Alvaro N\'{u}\~{n}ez, John Schliemann, and A. H. MacDonald}
\address{Department of Physics,
University of Texas, Austin, Texas 78712-1081}
\maketitle

\begin{abstract}
We present a simple model for the electron-phonon interactions between
the energy subbands in polyacene field-effect transistors and the vibrations
of the crystal.
We introduce a generalized Su-Schrieffer-Heeger model, arguing that
the strongest electron-phonon interactions in these systems originate from the
dependence of inter-molecule hopping amplitudes on collective molecular motion.
We compute the electron-phonon spectral function $\alpha^2F(\omega)$ as a
function of two-dimensional  hole density and the coupling strength constant.
Our results are in agreement with the sharp onset of
superconductivity near half-filling discovered in recent experiments 
by Sch\"on {\it et al.} \cite{Batlogg} and predict an increase of $T_c$ with pressure.
We further speculate on the implications that the observation of the quantum Hall effect
in these systems has on the effective band mass in the low carrier density regime.
\end{abstract}

\vskip2pc]

Present studies \cite{Batlogg,BatloggPRL}
of high-mobility organic transistors by Batlogg and collaborators
have shown an immense richness of tunnability 
of their electronic properties with gate voltage of in a single device. 
New possibilities for studying
the physics of coherent {\em band} quasiparticle transport in organic semiconductors
have been open by the high quality of these organic single crystals
and of their interfaces with AlO$_{3}$ dialectics.
To date, however,
the analysis of organic field-effect-transistor electronic systems has been hampered by the absence of
simple and reliable models for their electronic quasiparticles and for the interactions of
these quasiparticles with each other and with vibrations of the host lattice.  Indeed the discovery of
coherent band quasiparticle properties and of the quantum Hall effect in these
systems, which are usually thought of as being complex and relatively disordered,
has been one of the major surprises that has emerged from recent materials quality advances.

At temperatures above  $\approx 20 K$ the transport electronic properties 
of polyacene semiconductors are well described by
{\em small polaron} theory\cite{Kenkre}.
However, this small polaron theory is based on localized molecular orbitals
and is therefore unable to describe the low-temperature
{\em band} quasiparticle behavior seen in Battlogg {\em et al.'s} samples.
Schubnikov-de Haas transport measurements in these samples
demonstrate a quasiparticle mean-free-paths in excess of 1000 lattice constants,
demonstrating that
the low-temperature regime can be described only by starting at the opposite limit and building
a theory based on delocalized band quasiparticles states.  
Nevertheless, the interactions of quasiparticles
with lattice vibrations remain strong, and are presumably responsible for superconductivity.
We propose a simple 
tight-binding model for the quasiparticle bands originating from the $\pi$-molecular orbitals
and for the interaction of these bands with the host crystal vibrational excitations.
Our theory is parameter free once the band mass has been specified \cite{BatloggPRL}.

We focus, for simplicity, on the polyacene compound anthracene, 
which is monoclinic ($P2_{1/a}$) with two basis molecules whose orientations are related by a
gliding plane symmetry. 
We start by considering the field-induced energy bands obtained neglecting electron-phonon
interactions.  The key question we need to address is the number of 2D subbands that are occupied at a
particular 2D density, $n_{2D}$.  We find that all carriers reside in a single
subband up to much higher density in these transistors than in their inorganic counterparts,
principally because of the relatively small dialectic constants
($\epsilon\sim 3.5$) and the large effective in plane band mass ($m_\perp=1.5 m_e$) and
in spite of the rather large c-direction effective band mass $m_z\sim 3 m_e$ \cite{Batlogg}.
Our conclusion is based partially on envelope-function density-functional calculations
for which the local-density-approximation (LDA) can be used to estimate many-body
effects\cite{Ando_Fowler_Stern} that favor a single subband.
In our LDA calculations we find that the second subband
is first populated at $n_{2D} \approx 5-7\times 10^{14}{\rm cm}^{-2}$, a density just
larger than the highest achieved in experimental systems. 
Since the experimental signatures of second subband
occupation would be unambiguous and no anomalies in 2D density dependence have been reported,
we conclude that a single 2D subband is occupied up to the highest densities and that the
lowest subband is strongly localized in the top layer for $n_{2D}$ larger than $\approx 10^{13}
{\rm cm}^{-2}$.

Because Huckel model intra-molecular hopping parameters have no linear dependence
on low energy intra-molecular vibration normal coordinates, our model Hamiltonian
includes only inter-molecular hopping parameters \cite{lannoo}.
Then to model the electron-phonon interaction, we expand each $a-b$ plane hopping
parameter to first order in the twelve coordinates that describe rigid rotations and
displacements of neighboring polyacene molecules (as it is usually done in the
Su-Schrieffer-Heeger model \cite{SSH}):
\begin{eqnarray}
t&=&t_0+ \sum_{m=1}^6\sum_{\mu}\tilde{\bf t}_{\mu,m}
\tilde{u}_{\mu,m}
\label{hop_integral}
\end{eqnarray}
where $\tilde{u}_{\mu,m}$ is the generalized displacement coordinate,
$\mu =1,2$ is the molecular basis index, $m=1,2,3$ denote
displacements of the molecular center of mass along
$\hat x$,$\hat y$ and $\hat z$ directions, and $m=4,5,6$ denote
angular displacements around the $1,2,3$ principal axes
angular displacements around the $1,2,3$ principal axes
of each molecule. The electron-phonon interaction parameters
$\tilde{\bf t}_{\mu,m}\equiv \partial t /\partial \tilde{u}_{\mu,m}$
are calculated based on the assumption of
proportionality between hopping integrals and overlap integrals between HOMO or LUMO
orbitals on adjacent molecules.
These are calculated using standard
H\"{u}ckel approximation HOMO $\pi$ orbitals
and hence appropriate to the hole systems on which we focus.

We next present expressions for the interactions terms in the Hamiltonian between
phonons of the host crystal and the 2D band quasiparticles.
The phonon frequencies and polarization vectors are obtained
by solving the standard secular equation involving the dynamical matrix.
These eigenmodes are in general a mixture of displacements and rotations.
We compute the dynamical matrix following the procedure outlined in Ref. \cite{Taddei_Dorner}.
The phonon density-of-states and dispersion curves that emerge from these
calculations have been reported elsewhere and are omitted here for the sake of
brevity\cite{elsewhere}. 
The rigid-molecule vibration approximation used here is a convenient but inessential approximation that
can be complemented if necessary by including coupling to important isolated molecule vibrations;
it is however reasonably accurate for low-frequency vibrations of 
anthracene becoming less reliable for larger polyacenes.

Combining these ingredients we finally write the Hamiltonian (ignoring the
Coulomb interaction term) as
${\cal H}={\cal H}_{\rm 2D}^0+{\cal H}_{\rm 2De-vib}+{\cal H}_{\rm vib}$,
where
\begin{eqnarray}
{\cal H}_{\rm 2De-vib}&=&
\frac{1}{\sqrt{N}}\sum_{\kk\in{\rm BZ}}\sum_{\nu\alpha\beta,i,j}
\sum_{\QQ\in {\rm BZ}}
g_{i,\alpha;j,\beta}(\kk,\QQ,\nu)\times\nonumber\\&&
[\ah_{\QQ,\nu} +\ah^\dagger_{-\QQ,\nu}]\ccd_{i,\alpha [\kk+\qq]}
\cc_{j,\beta \kk}
\label{e-ph}
\end{eqnarray}
and
\begin{eqnarray}
&&g_{i,\alpha;j,\beta}(\kk,\qq,q_z,\nu)=
(-1)^i\frac{\tilde{f}_{\alpha,\beta}(q_z)}{2}
\sum_{\dd',\mu,m} \sqrt{\frac{\hbar}{2
\tilde{M}_{\mu,m} \omega_\nu(\QQ)}}\nonumber\\&&
\times\tilde{\bf t}_{\mu m}(\dd')\epsilon_{\mu m}(\QQ,\nu)
e^{i \qq\cdot\dd_\mu}(e^{-i\dd'\cdot[\kk+\qq]}
+(-1)^{(i-j)}e^{i\dd'\cdot\kk}).
\label{Tij}
\end{eqnarray}
In Eq.~\ref{Tij} $\tilde{f}_{\alpha,\beta}(q_z)=\sum_{z_l}a^{\alpha *}_l a^{\beta}_{l}
e^{iq_z z_l}$ is the form factor of the subband (indexed by $i,j$) 
tight-binding wave function and the brackets indicate reduction to the 2D BZ. In our
calculations we use the strict 2D limit
($\tilde{f}_{\alpha,\alpha}(q_z)=1$) because of the strong quantum
confinement found in  LDA and Hartree calculations.

\begin{figure}[h]
\epsfxsize=3.2in
\centerline{\epsffile{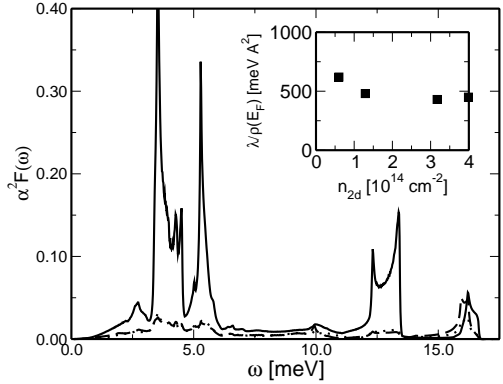}}
\caption{$\alpha^2F(\omega)$ for $n_{\rm 2d}=4.0 \times 10^{14}
\,{\rm cm}^{-2}$ (solid line),
$n_{\rm 2d}=1.4 \times 10^{14} \,{\rm cm}^{-2}$ (dotted line),
and $n_{\rm 2d}=6.7 \times 10^{13} \,{\rm cm}^{-2}$ (dashed line).}
\label{alpha2F}
\end{figure}

We next define the electron-phonon interaction spectral function as
\begin{eqnarray}
&\alpha&^2F(\omega)\equiv \frac{1}{N g(\mu)}
\sum_{\kk,\kk',i}\sum_{q_z,\nu} |g_{i,1;i,1}(\kk,[\kk'-\kk],q_z,\nu)|^2
\nonumber\\&&
\times\delta(\epsilon_i(\kk)-\mu) \delta(\epsilon_i(\kk')-\mu)
\delta(\omega-\omega_\nu([\kk'-\kk],q_z)
\end{eqnarray}
where $g(\mu)=\sum_\kk \delta(\epsilon_i(\kk)-\mu)$ is the electronic density of
states, and $[\kk'-\kk]$ denotes the projection of the $\QQ=(\kk'-\kk,q_z)$
vector into the three dimensional BZ.   The results of these calculations
for densities $n_{\rm 2d}=4.0 \times 10^{14}
\,{\rm cm}^{-2}$,
$n_{\rm 2D}=1.4 \times 10^{14} \,{\rm cm}^{-2}$,
and $n_{\rm 2D}=6.7 \times 10^{13} \,{\rm cm}^{-2}$ are
shown in Fig. \ref{alpha2F}. The peak locations in
the low-frequency rigid molecule vibration regime agree with those
observed in infrared absorption and tunneling data in pentacene crystals \cite{BatloggPRL}.
The mass enhancement factor $\lambda$, obtained from the integration of
$2F\alpha^2(\omega)/\omega$, close to half filling is $0.25$. 
Its sharp decay away from
half-filling (driven by the density of states) helps explain the sharp    
superconductivity onset observed in the experiments, and as a consequence
we expect a suppresion of superconductivity beyond half-filling from 
this simple theory.
We estimate the superconducting critical temperature by the
[dimension independent] BCS expression $k_B T_c\approx \hbar\omega_{D} \exp[-1/\lambda]$.
Such expression, using $\hbar\omega_{D}/k_B\approx 150 K$ from Ref. \cite{BatloggPRL}, yields $\sim 3 K$in agreement with the experiments\cite{Batlogg}.
We note that the inclusion of the second-nearest-neighbour hopping tends to push the
density of states peak to higher energies (densities). Adding a finite life time of the
quasiparticles, will broaden these peaks, hence creating a plateau
in the  n$_{\rm 2D}$-dependence of $T_c$ shifted to
the right of half filling as observed in the experiments \cite{Batlogg}.
This last effect would lower also the  $T_c$ calculated at $n=4.00 \times 10^{-12}{\rm cm}^{-2}$,
worsening experimental agreement.
However, given that $E_F$ does not satisfy $E_F>>\hbar\omega_{max}$, such estimates of $T_c$ from
$\lambda$ must be considered qualitative.

In summary we have presented a theory of the low-temperature quasiparticle
bands in polyacene field effect transistors, and of the electron-phonon coupling
with the host molecular crystal.  
Our calculations indicate that the quasiparticles lie in a
single 2D tight-binding band up to the highest densities that have been achieved
experimentally at present and that the most important electron-phonon interactions
arise from the influence of approximately rigid molecular translations and
rotations on hopping between HOMO and LUMO orbitals on adjacent molecules.
This picture implies a dependence of superconducting critical temperature on
molecular lattice constant which contrasts with the case of doped fullerene
superconductors.  For the fullerenes, the important electron-phonon interactions
are intra-molecular so that $T_c$ depends on inter-molecular hopping only
through the density-of-states.  Decreasing the lattice constant, increases
hopping, and hence decreases the density-of-states and $T_c$\cite{fullerences}.  
Here decreasing the lattice constant will also strengthen the important electron-phonon interactions.
According to our theory, the latter effect dominates and $T_c$ will increase with
decreasing lattice constant.  We find that phonon-mediated electron-electron
interactions in polyacene molecular crystals are strong only when the 2D tight-binding
band is close to half-filling and its density of states is relatively large.
This occurs for 2D densities comparable to $4 \times 10^{14} {\rm cm}^{-2}$, where
superconductivity turns on relatively abruptly.  At lower carrier densities,
$\lambda \ll 1$ and electron-electron interactions are dominated by repulsive
Coulomb interactions, consistent with the occurrence of the fractional quantum Hall effect.
We also note that the simple fact that the fractional quantum Hall effect is observed 
points to the distinct possibility that the band mass can be highly reduced
at those lower carrier densities since, otherwise, from rough estimates of the 
known experimental values at higher carrier densities, the $r_s$ value would be too
large to allow anything but a Wigner crystal ground state.

The authors acknowledge helpful discussions with P. Barbara,
B. Batlogg, A. Dodabalapur,  Y. Joglekar, T. Jungwirth, and P. Rossky.
This work was supported by the Deutsche
Forschungsgemeischaft, by the Welch Foundation and the by the National Science
Foundation under grant DMR0115947.

\end{document}